\documentclass[preprint2]{aastex}

\begin{document}

\title{Butterfly Diagram and Activity Cycles in HR 1099}

\author{Svetlana V. Berdyugina}
\affil{Institute of Astronomy, ETH Zurich, 8092 Zurich, Switzerland}
\affil{Tuorla Observatory, University of Turku, V\"ais\"al\"antie 20, FIN-21500, Piikki\"o, Finland}
\email{sveta@astro.phys.ethz.ch}

\and
\author{Gregory W. Henry}
\affil{Center of Excellence in Information Systems, 
       Tennessee State University, Nashville,  TN 37203}
\email{henry@schwab.tsuniv.edu}

\begin{abstract}
We analyze photometric data of the active RS CVn--type star HR~1099 for the 
years 1975--2006 with an inversion technique and reveal the nature of 
two activity cycles of 15--16 yr and 5.3$\pm$0.1 yr duration. The 16 yr 
cycle is related to variations of the total spot area and is coupled with 
the differential rotation, while the 5.3 yr cycle is caused by the symmetric 
redistribution of the spotted area between the opposite stellar hemispheres 
(flip-flop cycle).  We recover long-lived active regions comprising two 
active longitudes that migrate in the orbital reference frame with a variable 
rate because of the differential rotation along with changes in the mean 
spot latitudes.  The migration pattern is periodic with the 16 yr cycle. 
Combining the longitudinal migration of the active regions with a 
previously measured differential rotation law, we recover the first stellar 
butterfly diagram without an assumption about spot shapes.  
We find that mean latitudes of active regions at opposite longitudes change 
antisymmetrically in the course of the 16 yr cycle:  while one active region 
migrates to the pole, the other approaches the equator.  This suggests a 
precession of the global magnetic field with respect to the stellar rotational 
axis.
\end{abstract}


\section{Introduction}

HR~1099 (= V711~Tau) is the most extensively studied RS~CVn system, 
consisting 
of a K1 IV primary and a G5 IV secondary in a 2.8 day orbit.  The primary, 
being the faster rotator, is the first cool active star whose surface was 
mapped with the Doppler imaging (DI) technique \citep{vogt&pen83}.  It has 
a record 46 surface maps obtained between 1981 and the present 
\citep{don92,don03,don99,vogt99,str&bar00,gar03,pet04}.  Almost all images 
of the star feature a prominent, cool, polar cap along with a variety of 
high- and low-latitude spots. The area of the polar spot and its extensions 
may vary cyclically and correlate with the mean brightness of the system 
\citep{vogt99}. 

The photometric variability of HR~1099 was discovered in 1975 by 
\citet{lan&hal76} and has been followed regularly for more than 30 years.  
The data before 1993 were collected and analyzed by \citet{hehh95}.  They 
found periodicities of 5.5$\pm$0.3 and 16$\pm$1 yr, respectively, 
in the star's mean 
and maximum brightness, suggesting the presence of activity cycles.  Using 
a model with two circular spots, they identified 15 different dark spots 
within the 18 yr span of the data with lifetimes averaging between 1.5 
and 2.0 yr.  The spot rotation periods ranged from 2.8245 to 2.8412 days 
and bracketed the orbital period by $-$0.47\% to +0.12\% days, implying 
significant differential rotation.  \citet{vogt99} measured the differential 
rotation from DI and found that it is of opposite sign and about a factor of 
56 less than for the Sun.  With newer stellar images, however, \citet{pet04} 
argued that the differential rotation law for HR~1099 is of the same sign 
as the Sun and a factor of 2 larger than determined by \citet{vogt99}.  
The range of polar and equatorial periods determined by \citet{pet04} 
encompasses all periods derived from the photometry of \citet{hehh95}.

\citet{moh&rav93} interpreted the photometric variability of HR~1099 as 
the result of two active longitude regions migrating in the orbital reference 
frame with constant but very different rates.  Two active longitudes (ALs) 
separated by about 180\degr\ and locked in synchronism with the orbital 
period have been surmised by \citet{hehh95}.  In comparison with other 
RS~CVn--type stars, however, the ALs on HR~1099 appeared very broad, 
110\degr--140\degr, implying smearing by differential rotation.

In the present Letter we analyze all photometric data collected up to date 
for HR~1099, previously published and newly obtained.  Employing a technique 
based on inversions of stellar light curves into images of the spot filling 
factor \citep{ber02}, we recover locations of active regions and search for 
long-lived ALs.  Taking into account the effect of differential rotation, 
we are able to recover a stellar butterfly diagram, which, in contrast to 
previous studies \citep[e.g.,][]{kat03}, does not involve assumptions about 
spot distributions, shapes, and numbers.

\section{Observations}\label{sec:obs}

We have analyzed the Johnson $V$ photometric observations collected earlier 
from the literature and from the T1 0.25~m and T3 0.40~m automatic photometric 
telescopes (APTs) by \citet{hehh95}, that span the years 1975--1993.  We 
have also included new observations for the years 1994--2006, acquired with 
the T3 APT.  The T1 and T3 APTs, along with several others, are housed at 
Fairborn Observatory in southern Arizona.  Further details on the operation 
of the APTs and the reduction of the data can be found in \citet{h95a,h95b} 
and \citet{ehf03}.  The complete set of Johnson $V$ differential magnitudes 
is listed in Table~\ref{tab:phot} and plotted in Fig.~\ref{fig:alon}.  The 
differential magnitudes have been converted to apparent magnitudes in 
Fig.~\ref{fig:alon}$a$ with a $V$ magnitude of 4.28 for the comparison star 
10~Tau (= HR~1101). The faint, close, visual companion of HR~1099 was always 
included in the photometer diaphragm and so is included in the light of 
HR~1099.

The $V$ observations were divided into 109 subsets
within which the light curve was observed to be stable over each
time interval.   The average length of each subinterval was $\leq$40 days.
Julian Dates of the observations were converted to orbital phases with the 
ephemeris
\begin{equation}
T_{\rm conj}=2,442,766.08+2.83774E
\label{eph}
\end{equation}
adopted from \citet{fek83}, where zero phase corresponds to the conjunction 
with the K1 star closest to the Earth. 
The error in the period of 10$^{-5}$\,days results in a phase uncertainty 
of only 0.04 after 30 years.
According to the Fourier analysis of \citet{hehh95}, the star exhibits the 
presence of a small ellipticity effect with a full amplitude of 0.016 mag in 
$V$. This effect was removed from each data set before the light-curve 
inversions were performed.

\section{Activity cycles and active longitudes}\label{sec:cyc}

\begin{figure}
\plotone{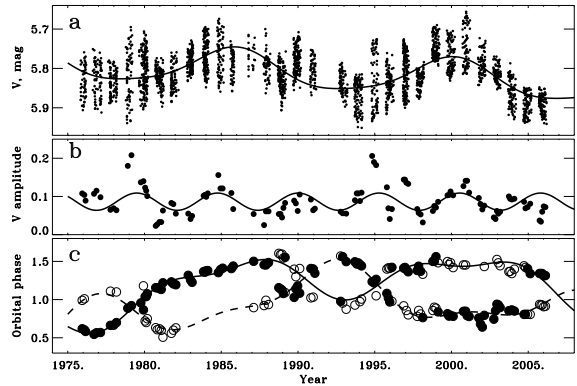}
\caption{Activity patterns on HR~1099.
         ($a$) All $V$-magnitude observations analyzed in this Letter. 
The data before 1993 are from \citet{hehh95}.  Cyclic variations of the 
mean brightness are revealed with a harmonic fit ({\emph solid curve}) 
with a period of $\sim$15 years.
         ($b$) Peak-to-peak amplitude of the light curves 
({\emph filled circles}).  A harmonic fit ({\emph solid curve}) reveals 
the period of 5.3$\pm$0.1 yr. 
         ($c$) Phases of the spots recovered with the light-curve
inversions.  Filled circles denote the larger or single active region
observed at a given time, while open circles indicate a second, smaller
region if observed.  Migration paths of the ALs are
indicated with solid and dashed curves, which are harmonic fits to
the measurements with a long-term trend.  The migration pattern is
apparently repeated in about 16 years.
}
\label{fig:alon}
\end{figure}

For each of the 109 light curves, we measure the maximum, mean, and minimum 
$V$ magnitudes as well as the peak-to-peak amplitude in $V$.  Period analysis 
of these values confirms the cycles found by \citet{hehh95}.  The mean 
brightness variations show a cycle of $\sim$15 years (Fig.~\ref{fig:alon}$a$), 
and similar cycle lengths of 15--16 years are seen in the maximum and minimum 
magnitudes. 
The peak-to-peak amplitude in $V$ varies with a cycle of 5.3$\pm$0.1 yr 
(Fig.~\ref{fig:alon}$b$).  This cycle also has an influence on variations 
in the maximum, mean, and minimum $V$ magnitudes.  Similarly, the 
16 yr cycle also influences the amplitude variations, as the largest 
amplitude of 0.2 mag was observed at the maxima of the spottedness in 
1979 and 1994.  

\begin{figure}
\plotone{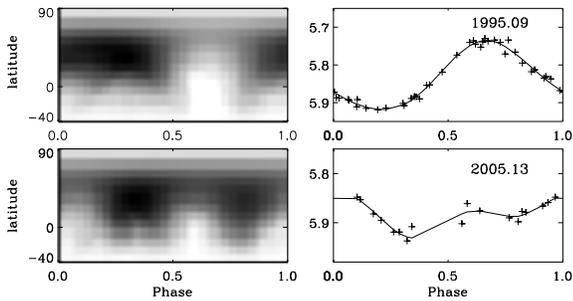}
\caption{Examples of the spot filling factor distribution on HR~1099 
({\emph left}) and corresponding light curves ({\emph right}).  
The stellar surface was modeled 
with a grid of 10$^{\circ}\times$10$^{\circ}$.  The spot filling factor at 
each grid point indicates relative contributions from two temperature 
components, the hot photosphere and cool spots.  In the images, the darker 
area implies a higher spot filling factor.  Observed light curves are shown 
as crosses and the fit as the solid line.  In contrast to direct modeling 
of light curves, the inversions do not involve any assumptions about spot 
shapes, numbers, or latitudes.
}
\label{fig:image}
\end{figure}

To determine the large-scale spot distribution on HR~1099, we apply an 
inversion technique to the light curves using a two-temperature approximation 
\citep{ber02}.  
The inversion of the light curve results in the distribution of the spot 
filling factor over the stellar surface, i.e.,\ a stellar image 
(Fig.~\ref{fig:image}). 
We adopted values of 4700 and 3500\,K for the photosphere and spot 
temperatures, respectively, following \citet{hat96}.  Also, we assumed 
that the star is spotless at $V=5.6$ mag, which is about 0.05 mag brighter 
than the brightest magnitude observed for HR~1099.  The inclination of the 
rotational axis to the line of sight is chosen to be 40$^\circ$; 
inclinations in the range of 33\degr--45\degr\ have been used for 
the previous DI mentioned above.

From the stellar images, we recover spot longitudes with the maximum spot 
filling factor 
\citep[see details in][]{ber02}. The accuracy of the spot longitudes
depends on the broadness of the light curve minimum and is on average 0.05 
in phase.
When two minima are seen in the light curves, two spot concentrations are 
recovered in the images. The time dependence of the spot phases is shown in 
Fig.~\ref{fig:alon}$c$ (the data are available on request). Our derived 
spot phases constitute two sequences that represent two migrating active 
longitude regions separated on average by 180$^\circ$, similar to 
the previous studies of HR~1099 \citep{moh&rav93,hehh95,lan06} and to
other RS~CVn--type stars \citep{ber&tuo98,fek&hen05}. 
\cite{lan06} related the long-term trend in the migration of the ALs 
with binary orbital period variations.

We notice that the two active regions regularly approach each other and 
merge into a large single region.  Subsequently, the single region divides 
back into two spot regions.  Thus, we are seeing the crossing of the migration 
paths of the two ALs. 
In the presence of differential rotation, such behavior suggests 
variations in mean spot latitudes.
Three intersections were clearly observed in 1979, 1990, and 1995.  At the 
same times, the amplitude of the brightness variations reached a maximum 
(Fig.~\ref{fig:alon}$b$), which implies a rearrangement of spots onto one 
stellar hemisphere and supports the idea of merging active regions and 
crossing ALs.  
The intersections in 1979 and 1990 are also seen in the reconstruction 
by \cite{lan06}.  However, they missed the one in 1995 due to lack of data. 
The shorter ($\sim$5 yr) and longer ($\sim$11 yr) time intervals between 
the intersections comprise the 16 yr cycle.

In Fig.~\ref{fig:alon}$c$, the two migration paths are plotted with 
two fits to 
the observed phases.  The fits are obtained with a two-harmonic Fourier 
series and a long-term trend, which together are able to reproduce the 
nonsymmetric shapes of the curves.  
The assignment of the active regions to one or the other AL has been 
done under the assumption that the active region continues its migration 
with the same rate after the intersection of the ALs, which is natural in 
the presence of differential rotation.
Although the fits have been calculated separately for the two ALs, both 
reveal similar shapes that are mirrored with respect to each other in 
phase.  The migration patterns appear to be periodic with a 16$\pm$3 
yr cycle.  
The standard deviation of the fits is as small as 0.06 phase units.  An 
extrapolation of the fits suggests that the next crossing of the active 
longitudes will occur in 2007.  Observation of the ALs crossing in 2007 
with an increase in photometric amplitude will be a good test for the 
present model.

In addition to the migration patterns and their intersections, abrupt 
switches of the dominant activity between the ALs occurred in 1988, 
1992--1993, 
1998--1999, and 2005, all at the times when the amplitude of the light curve 
was near the minimum.  Such events periodically occur in RS~CVn-- and 
FK~Com--type stars and are known as flip-flop cycles \citep{ber&tuo98,kor02}.  
In HR~1099, the flip-flops occurred when the active regions at opposite 
longitudes became about the same size; that reduced the amplitude of the 
light curve, and the dominate activity switched to the opposite longitude.  
Thus, flip-flops in HR~1099 are apparently governed by the 5.3 yr cycle but 
appear to be disturbed by the differential rotation and intersections of the 
active longitudes. 

\section{The butterfly diagram}\label{sec:fly}

\begin{figure}
\plotone{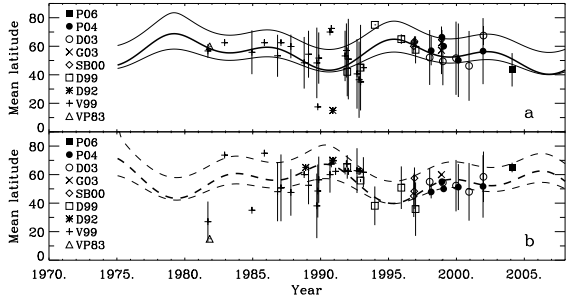}
\caption{Butterfly diagram of HR~1099.
         ($a$) Mean latitudes recovered from the fit to the active longitude 
             shown in Fig.~\ref{fig:alon}$c$ as the solid curve.
         ($b$) Mean latitudes recovered from the fit to the active longitude 
             shown in Fig.~\ref{fig:alon}$c$ as the dashed curve.
         In both plots, the mean latitudes of spots in the opposite hemispheres
         measured from DI are shown with different symbols
         corresponding to the following sources: 
VP83 \citep{vogt&pen83}, 
V99  \citep{vogt99}, 
D92  \citep{don92}, 
D99  \citep{don99}, 
SB00 \citep{str&bar00},
G03  \citep{gar03}, 
D03  \citep{don03},
P04  \citep{pet04}, and
P06  (P.~Petit 2006, private communication).
Standard deviations of the mean DI latitudes (if several spots were 
observed) are shown as vertical dashes.  The thick curves are the solutions 
using the best parameters of the differential rotation law by \cite{pet04}. 
The thin curves indicate the $\pm$1\,$\sigma$ interval for the best parameters 
obtained in this Letter (filled circle in Fig.~\ref{fig:difrot}).
The two solutions are statistically equivalent.
}
\label{fig:butterfly}
\end{figure}

The shape of the spot migration curves for HR~1099 is reminiscent of the 
solar ALs \citep{ber&us03}.  On the Sun, such a migration pattern is 
determined 
by changes of the mean spot latitudes along with the surface differential 
rotation.  Assuming that a similar relation holds for HR~1099, we can recover 
mean spot latitudes on its surface, i.e., create butterfly diagrams for the 
active longitudes in the stellar hemisphere visible above the stellar 
equator.  For this purpose we use the differential rotation law:
\begin{equation}
\Omega=\Omega_{\rm eq} - \Delta\Omega\sin^2\psi,
\label{eq:dif}
\end{equation}
with the parameters estimated by \citet{pet04} from DI of HR~1099 for 
the year 2002, having the smallest variances: 
$\Omega_{\rm eq}$=2.2241$\pm$0.0004\,rad\,day$^{-1}$ and 
$\Delta\Omega$=15.2$\pm$0.8\,mrad\,day$^{-1}$.
By observing a phase shift of the ALs during a given time interval, we can 
determine an instant angular velocity $\Omega$, and, using Eq.~(\ref{eq:dif}), 
we can find the corresponding latitude $\psi$.  To increase accuracy, we 
determine phase shifts from the harmonic fits to the migration paths.  The 
results are shown as solid and dashed lines in 
Fig.~\ref{fig:butterfly}$a$~\&~$b$.  As expected, mean spot latitudes 
vary with the 16 yr cycle for both active longitudes.

\begin{figure}
\plotone{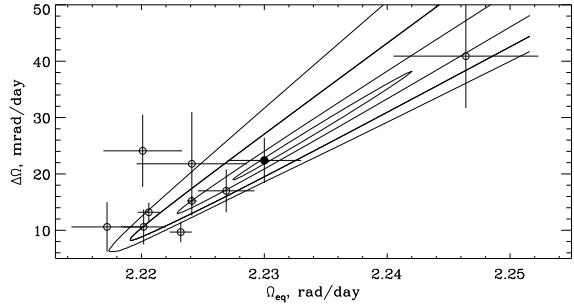}
\caption{The $\chi^2$ surface contours for the two parameters in 
Eq.~(\ref{eq:dif}) calculated from the AL migration and the DI mean spot 
latitudes.  The $\chi^2$=1 level is emphasized by a thick contour. The 
best-fit solution is indicated by the filled circle.  The measurements 
by \cite{pet04} are shown as open circles.  The strongly elongated shape 
of the minimum implies that the two parameters are almost linearly dependent. 
The distribution of the measurements along the major axis of the error 
ellipsoid suggests that they reflect errors in DI spot latitudes rather 
than true variations of the differential rotation.
}
\label{fig:difrot}
\end{figure}

The extent and shape of the butterfly diagram recovered from the AL 
migration can be compared with the measurements of spot latitudes from DI.  
From the 46 published images, we measured latitude extents of the recovered 
spots and calculated weighted mean latitudes for each of the two opposite 
ALs.  These are shown in Fig.~\ref{fig:butterfly}$a$~\&~$b$. The mean DI 
latitudes and the curves obtained from the AL migration are in good 
agreement, thus supporting our interpretation of the photometry.  
To estimate a possible range of the parameters in Eq.~(\ref{eq:dif}), 
we searched for the best fit (via $\chi^2$ minimization) to the mean DI 
latitudes combined with the angular velocity $\Omega$ measured from the 
AL migration. The result is shown in Fig.~\ref{fig:difrot}. It reveals a 
very elongated error ellipsoid encompassing measurements by \citet{pet04} 
for different years.  
Our best-fit parameters, $\Omega_{\rm eq}$=2.230$\pm$0.003\,rad\,day$^{-1}$ 
and $\Delta\Omega$=22$\pm$4\,mrad\,day$^{-1}$, indicate a somewhat stronger 
differential rotation on HR~1099 than the best parameters by \citet{pet04}, 
although the two measurements appear to be statistically equivalent.

The shape of the butterfly diagram for HR~1099 is reminiscent of the Sun's.  
The direction of the latitude migration, however, is different for the two 
ALs.  For the active region shown in Fig.~\ref{fig:alon}c with a solid line, 
preferred spot appearances first move relatively rapidly (for 5 years) to 
higher latitudes, up to 70$^\circ$, and then slowly (for 11 years) move back 
to mid latitudes, down to 40$^\circ$ (Fig.~\ref{fig:butterfly}$a$).  For the 
other active region (shown in Fig.~\ref{fig:alon}$c$ with a dashed line), 
the behavior of the spot latitudes is just the opposite. 
The latitude shifts are therefore always opposite for the two ALs.  
During 1982--1987, both ALs were at about the same latitude, and the 
migration rate remained constant.  In 1988--1993, one active longitude was 
occupied by large high-latitude (polar) spots, while low-latitude spots 
appeared around the opposite active longitude.  After 1997, the dominant 
active regions again moved to about the same latitude, thus completing the 
16 yr cycle. The opposite latitudinal migration of the two long-lived 
regions could be responsible for the contradictory measurements of the 
differential rotation by \citet{vogt99} and \citet{pet04}.

\section{Discussion}\label{sec:dis}

As noted by \citet{hehh95}, if the 16 yr cycle is akin to the 11 yr sunspot 
cycle, one might anticipate other signatures of such a cycle.  Indeed, the 
16 and 5.3 yr activity cycles in HR~1099 are similar to the sunspot cycles 
of 11 and 3.7 yr \citep{ber&us03} and the cycles observed on young solar 
analogs \citep{ber&jar05}.  The longer cycle in HR~1099 reflects the variation 
of the total spottedness and is coupled with the differential rotation.  
This cycle may be associated with an oscillating axisymmetric dynamo mode 
in which global polarity reversals occur in cycles of 22 years on the Sun or 
32 years on HR~1099.  No reversals have yet been observed on HR~1099 despite 
12 years of spectropolarimetric observations \citep{don03}.  Comparing the 
butterfly diagrams of the Sun and HR~1099, we suppose that a global polarity 
reversal on HR~1099 could have happened around 1990, just before the regular 
spectropolarimetric observations by Donati et al.\ began.  The next one can 
be expected in 2007.


The shorter activity cycle on HR~1099 is related to the flip-flop phenomenon; 
i.e., it reflects symmetrical redistribution of the spotted area between 
opposite stellar hemispheres.  In the light curves, the flip-flops are seen 
as a strengthening of the secondary minima and a diminishing of the primary 
one.  In the Doppler images of HR~1099 obtained by \citet{don99} and 
\citet{don03}, the flip-flop in 1992--1993 is seen as an abrupt change of the 
phase towards which the polar spot is tilted by about half a rotational 
period, while the flip-flop in 1998--1999 appears as a progressive shift of 
the polar spot extension to the opposite longitude.  The flip-flop cycles 
can be associated with a non-axisymmetric dynamo mode, which gives rise to 
ALs \citep{fluri&berd04}.  Similar to the Sun and young solar analogs, the 
flip-flop cycle on HR~1099 is about one-third of the length of the cycle 
regulating the total spottedness, which implies a certain relation 
between the non- and axisymmetric dynamo modes.

The fact that the butterfly diagrams for the two ALs are antisymmetric, i.e., 
that latitude shifts are always opposite in the opposite stellar hemispheres, 
implies that the axis perpendicular to the line connecting the ALs changes 
its direction with respect to the rotational axis by about $\pm$15\degr\ 
during the 16 yr cycle.  Thus, the two axes are separated the most in 
1990--1991 and 1995--1996 and aligned in 1982--1987 and 1998--2003, 
suggesting the 
precession of the nonaxisymmetric dynamo mode during the 16 yr cycle.  
Such a precession of the magnetic field was also suspected in the 
Zeeman-Doppler images of HR~1099 obtained by \citet{don03}, who noticed 
that the rings of the azimuthal field got progressively better centered in 
2000--2002, while the fraction of the magnetic energy attributed to the 
meridional field component was regularly increasing in 1992--1997 and 
decreasing starting from 1998 up to now.  They assumed that such behavior 
suggests that the main axis of symmetry of the azimuthal field component is 
varying in orientation with time on the timescale of at least one decade, 
which is in agreement with our findings.

Our results suggest that a few interesting events in the magnetic activity of 
HR~1099 may happen in 2007.  For instance, an extrapolation of the fits to 
the ALs predicts their next crossing to happen during this time.  Also, we 
can expect the mean spot latitude of the dominant active region to be 
different again and the rings of the azimuthal field to be decentered.  In 
addition, a global polarity reversal may finally be observed. The detection 
of such events would be of great importance for understanding stellar magnetic 
cycles and their underlying dynamo processes.

We thank Lou Boyd at Fairborn for making our automated photometry possible.
This research at Tennessee State University was supported by NASA grant
NCC5-511 and NSF grant HRD-9706268. SVB acknowledges the EURYI (European Young
Investigator) Award provided by the European Science Foundation 
(see www.esf.org/euryi).


\begin{deluxetable}{ccc}
\tablewidth{0pt}
\tablecaption{HR~1099 Photometric Data.
\label{tab:phot}}
\tablehead{
\colhead{HJD} & \colhead{$\Delta$V} }
\startdata
2,442,719.8650&  1.600\\
2,442,720.9300&  1.510\\
2,442,721.9050&  1.570\\
\enddata
\tablenotetext{*}{Table 1 is published in its entirety in the electronic 
edition of the {\em Astrophysical Journal}. A portion is shown here for 
guidance regarding its form and content.}
\end{deluxetable}

\end{document}